\documentclass[twocolumn,preprintnumbers,amsmath,amssymb,floatfix,superscriptaddress,nofootinbib]{revtex4-1}

\usepackage{amsmath,hyperref,amssymb}
\usepackage{graphicx}
\usepackage{dcolumn}
\usepackage{bm}
\usepackage{verbatim}
\usepackage{slashed}
\usepackage{cancel}
\usepackage{float}
\usepackage{enumerate}
\usepackage{bm}
\usepackage{setspace}
\usepackage{multirow}
\usepackage{hyperref}

\usepackage{enumitem}
\setlist{nolistsep}

\usepackage{latexsym}
\usepackage{amssymb}
\usepackage{epsfig,amsmath,graphics}
\usepackage{listings}
\usepackage{color}
\usepackage{dcolumn}
\usepackage{slashed}
\usepackage{cancel}
\usepackage{comment,latexsym} 
\usepackage{bm}
\usepackage{verbatim}
\usepackage{tabularx}
\usepackage{dcolumn}
\usepackage{color}

\def\be{\begin{equation}}
\def\ee{\end{equation}}
\def\bea{\begin{eqnarray}}
\def\eea{\end{eqnarray}}
\newcommand{\eref}[1]{Eq.\!~(\ref{#1})}

\newcommand{\nbsb}{\bar{n}\cdot \bar{\sigma}} 
\newcommand{\alc}{\mathcal{A}} 
 
\newcommand{\nsb}{n\cdot \bar{\sigma}} 
\newcommand{\nbs}{\bar{n}\cdot \sigma} 
\newcommand{\ns}{n\cdot \sigma}

\newcommand{\nbp}{\bar{n}\cdot \partial} 
\newcommand{\np}{n\cdot \partial}

\definecolor{colorTC}{rgb}{.2,.7,.2}

\begin{document}

\vspace*{-30mm}

\title{Collinear Superspace}
\author{Timothy Cohen}
\affiliation{\,Institute of Theoretical Science, University of Oregon, Eugene, OR 97403 \vspace{2pt}}
\author{Gilly Elor}
\affiliation{\,Center for Theoretical Physics, Massachusetts Institute of Technology, Cambridge, MA 02139\vspace{2pt}}
\author{Andrew J.~Larkoski\,}
\affiliation{\,Center for the Fundamental Laws of Nature, Harvard University, Cambridge, MA 02138 \vspace{2pt}}

\begin{abstract}
\begin{centering}
{\bf Abstract}\\[4pt]
\end{centering}
This letter provides a superfield based approach to constructing a collinear slice of $\mathcal{N}=1$ superspace.  The strategy is analogous to integrating out anti-collinear fermionic degrees-of-freedom as was developed in the context of soft-collinear effective theory.  The resulting Lagrangian can be understood as an integral over collinear superspace, where half the supercoordinates have been integrated out.  The application to $\mathcal{N}=1$ super Yang-Mills is presented.  Collinear superspace provides the foundation for future explorations of supersymmetric soft-collinear effective theory.

\end{abstract}

\vspace*{1cm}

\maketitle

\begin{spacing}{1.1}
Supersymmetry (SUSY) is a powerful framework for exploring the properties of quantum field theory.  There are many examples of extraordinary results derived for SUSY models, for instance the exact NSVZ $\beta$-function~\cite{Novikov:1983uc}, Seiberg duality~\cite{Seiberg:1994pq}, Seiberg-Witten theory~\cite{Seiberg:1994aj,Seiberg:1994rs}, and the finiteness of $\mathcal{N} = 4$ SUSY Yang-Mills (SYM)~\cite{Mandelstam:1982cb}.  Identifying models that manifest SUSY in non-trivial ways has yielded many fruitful developments, see~\cite{Komargodski:2009rz, Festuccia:2011ws, Kallosh:2016hcm, Ferrara:2016een, Dall'Agata:2016yof} for recent examples.  In this letter, we explore a new class of $\mathcal{N}=1$ SUSY effective field theory (EFT) models which live on a ``collinear slice" of superspace; defining this collinear superspace is the subject of this letter. 

The connection between collinear superspace and gauge theories becomes apparent in the infrared (IR), where the physics can be largely inferred from the presence of soft and collinear divergences. There is a rich history associated with the IR structure of gauge theories.  For example, a correspondence between the coefficients of Sudakov logs in Yang-Mills theory and the cusp anomalous dimension of Wilson loops was discovered as early as 1980~\cite{Polyakov:1980ca}.  The importance of these IR effects helped lead to the discovery of Soft-Collinear Effective Theory (SCET)~\cite{Bauer:2000ew, Bauer:2000yr, Bauer:2001ct, Bauer:2001yt, Bauer:2002uv, Hill:2002vw, Chay:2002vy, Beneke:2002ph}, which is a powerful formalism developed for resumming the IR divergences occurring for processes that are dominated by soft (low momentum) and collinear degrees of freedom; see~\cite{Becher:2014oda, StewartNotes} for reviews.  There exists an ever growing literature exploring practical applications of SCET to heavy meson decays~\cite{Bauer:2001cu,Chay:2002vy,Beneke:2002ph,Lunghi:2002ju}, LHC collisions~\cite{Bauer:2002nz,Stewart:2009yx,Mantry:2009qz,Becher:2010tm,Beneke:2010da}, and even WIMP dark matter systems~\cite{Baumgart:2014vma, Bauer:2014ula, Ovanesyan:2014fwa}.  Our purpose here is to lay the groundwork for supersymmetrizing SCET, in hopes of further illuminating non-trivial aspects of field theory.

SCET can be understood in terms of a mode expansion, where a power-counting parameter $\lambda$ is used to separate degrees-of-freedom that are ``near" a light-like direction, thereby capturing the IR dynamics as an expansion in $\lambda$, from the ``far" modes.  Integrating out these ``anti-collinear" degrees-of-freedom yields the effective Lagrangian of SCET.  Note that this procedure obscures the underlying Lorentz invariance of the theory, leaving behind the constraints known as reparameterization invariance (RPI) \cite{Marcantonini:2008qn}.  Given its spacetime nature, it is unclear that SUSY can be preserved in any meaningful way.  Our main result is to show how collinear superspace packages a SCET Lagrangian in a language that makes the SUSY of the EFT manifest.

To derive the collinear limit for a fermion requires integrating out the anti-collinear modes, which in practice are half of the full theory fermion helicity degrees of freedom (the momenta of the EFT fields are also constrained).  This procedure guides the construction here:  the EFT can be characterized in terms of half the supercharges for $\mathcal{N}=1$ SUSY -- the other half of the supersymmetries are non-linearly realized. We refer to this as ``integrating out" half of superspace, which leave behind a collinear subsurface of superspace.  Our procedure for deriving collinear superspace, which should be generally applicable to a wide class of SUSY EFTs, can be described by the following algorithm: 

\begin{center}
\noindent {\textit{General Algorithm}} 
\end{center}
\vspace{-5pt}
\begin{itemize}[leftmargin=*]
\item Find projection operators that separate the superfield into collinear/anti-collinear superfields~[\emph{e.g.}~\eref{eq:SuperfieldProjections}].
\item Starting with the superspace action for the full theory, integrate out the entire anti-collinear superfield. This will yield a constraint equation [\emph{e.g.}~\eref{eq:VectorSuperfieldConstraint}].
\item Based on the constraint equation, guess an ansatz for the equation of motion for the anti-collinear superfield in terms of collinear degrees-of-freedom~[\emph{e.g.}~\eref{eq:VectorAnsatz}]. 
\item Plug the ansatz into the full theory action to yield the superspace action of the effective theory~[\emph{e.g.}~\eref{eq:VectorSuperspaceLagrangian}].
\end{itemize}
\vspace{5pt}
In what follows, we will apply this procedure to the explicit case of $\mathcal{N}=1$ SYM. 

To begin, we will provide some conventions.  The SUSY EFT is defined in Minkowski space with signature $g^{\mu \nu} = \textrm{diag}\left(+1,-1,-1,-1\right)$.  The collinear direction is taken along the $\hat z$ light-cone direction: $n^\mu = \left(1,0,0,1\right)$.  The anti-collinear direction is defined by $n^2 = 0 = \bar{n}^2$ and $n \cdot \bar{n} = 2$. It is usually convenient to make the explicit choice $\bar{n}^\mu =\left(1,0,0,-1\right)$.  Four vectors are expanded as $p^\mu = (n\cdot p,\, \bar{n} \cdot p , \,\vec{p}_\perp)$, where ``$\perp$" refers to the two directions perpendicular to both $n$ and $\bar{n}$. A state is collinear to the light-cone when it lives within a momentum shell which scales as $p_{n}^{\mu} \sim  (\lambda^2,1, \lambda)$, where $\lambda \ll 1$ is the SCET power counting parameter.  The virtuality for the collinear modes in the effective theory $p^2 \sim \lambda^2$ can be interpreted as closeness to the light cone.  Similarly, an anti-collinear momenta scales as $p_{\bar{n}}^{\mu} \sim (1, \lambda^2, \lambda)$.  Fields also scale as powers of $\lambda$; the power counting rules can be inferred from the appropriate kinetic terms, and must be necessarily tracked when determining the order of a given operator.

As discussed previously, studying the collinear fermion EFT will provide insight for the derivation of collinear superspace.  A two-component left-handed Weyl spinor can be decomposed into collinear and anti-collinear momentum modes using projection operators; $u~\!=~\!\!\left(P_n + P_{\bar{n}}\right) u = u_{n} + u_{\bar{n}}$, where
\begin{align}
P_n =  \frac{\ns}{2}\frac{\nbsb}{2} \quad\quad \quad P_{\bar{n}}  = \frac{\nbs}{2} \frac{\nsb}{2} .
\label{eq:projectionOps}
\end{align}
These also correspond to chiral projection operators that distinguish the fermion's spin states in the collinear limit~(a detailed discussion of two-component collinear fermions will be given in a forthcoming paper~\cite{SUSYSCET}). The anti-collinear modes $u_{\bar{n}}$, which scale as $\mathcal{O}(\lambda^2)$, are power suppressed relative to the collinear ones $u_{n} \sim \mathcal{O}(\lambda)$.  Therefore, $u_{\bar{n}}$ should be integrated out using the classical equation of motion:
\bea
u_{\bar{n}} =   -\frac{\nbs}{2}  \frac{1}{\bar{n} \cdot \mathcal{D}}\, \big(\bar{\sigma} \cdot \mathcal{D}_\perp\big) \,u_n ,
\label{eq:anticollEOM}
\eea
yielding the following Lagrangian for a charged collinear fermion
 \bea
\!\! \mathcal{L}_u = i \,u_n^{\dagger} \left( \, n \cdot \mathcal{D} +  \,\bar{\sigma} \cdot \mathcal{D}_{\perp} \frac{1}{ \,\bar{n}\cdot \mathcal{D}} \, \sigma \cdot \mathcal{D}_{\perp} \right) \frac{\bar{n}\cdot \bar{\sigma}}{2}\,  u_n\,,
\label{eq:LSCET}
\eea
where $\mathcal{D}$ is the covariant derivative appropriately power expanded when acting on collinear fields, and the non-local operator is defined in terms of its momentum eigenvalues, see~\emph{e.g.}~\cite{StewartNotes, Becher:2014oda}. 

The gauge bosons of the full theory can simply be expanded as $A^\mu = A_n^\mu + A_{\bar{n}}^\mu$, with a corresponding gauge Lagrangian for each sector $\mathcal{L} = - \frac{1}{4}\left(F_n^{\mu \nu} \right)^2 - \frac{1}{4}\left(F_{\bar{n}}^{\mu \nu} \right)^2$. Note that the gauge field is decomposed into components that scale as momentum along the collinear, anti-collinear, and perpendicular directions. Thus the field strength $i\, g\, F_n^{\mu \nu} = \bigl[ \mathcal{D}^\mu, \mathcal{D}^\nu \bigr]$ scales inhomogeneously with $\lambda$.  However, after contractions the gauge boson Lagrangian density does scale homogeneously: $F^2 \sim \lambda^4$. In what follows, we focus on the collinear modes, as the soft modes can be decoupled at leading power by a field redefinition \cite{StewartNotes,Becher:2014oda}. 

Collinear superspace is on-shell, \emph{i.e.}, only physical degrees-of-freedom will be present in the Lagrangian.  To this end, it is convenient to work in Light Cone Gauge (LCG) which corresponds to the non-(space-time)-covariant gauge choice $\bar{n}~\!\!\cdot\!\!~A = 0$, see \emph{e.g.}~\cite{Leibbrandt:1987qv} for a review.  Additionally, the mode $n\!\!~\cdot\!\!~A$ is non-propagating in this gauge (with respect to light-cone time) -- it can be integrated out by solving the classical equation of motion.
The two remaining bosonic physical degrees of freedom, the transverse components of the gauge field, can be recast as a complex scalar $\alc$, defined by
\bea
\partial_\perp \cdot A_{n \perp} \equiv - \partial^* \! \alc - \partial  \alc^*,
\eea
where $\partial$ and $\partial^*$ are also implicitly defined by this equation~\cite{Leibbrandt:1987qv}.  Then $\mathcal{L} = - \frac{1}{4} F^2_n \rightarrow \alc^* \Box \alc + \mathcal{L}_\text{int}$. 

For concreteness, our focus here is on-shell $\mathcal{N}=1$ SYM.  The fermionic degree of freedom $u_n$ (the single remaining spin state in the EFT after the SCET projection) is the collinear gaugino whose superpartner is the bosonic light cone scalar $\alc$. In~\cite{SUSYSCET}, we will provide a detailed derivation of the corresponding collinear SCET Lagrangian along with a demonstration that it passes checks necessary for EFT consistency, \emph{e.g.} RPI.

The $\mathcal{N} = 1$ supercharges are defined by the graded algebra
$\big\{
Q_\alpha,Q^\dagger_{\dot \alpha}
\big\} = 2\,\sigma^\mu_{\alpha\dot\alpha}\, P_\mu$, 
where the spinor and anti-spinor indices run over $\alpha, \dot{\alpha} = 1,2$. Power counting the generator of translations $P_\mu = i \,\partial_\mu$ as appropriate for collinear momenta, yields the scaling of the algebra in the EFT:
\begin{eqnarray}
 \!\! \!\!  \!\! \!\!\!\Big\{Q_\alpha, Q_{\dot{\alpha}}^\dagger \Big\}  = 2\,i
\left[ \!
\begin{array}{cc}
\np & \sqrt{2}\,\partial^*  \\ 
\sqrt{2}\, \partial & \nbp  \end{array}\! \right] _{\alpha \dot{\alpha}} 
\!\!\sim \left[\! \begin{array}{cc}
\mathcal{O}(\lambda^2) &\mathcal{O}(\lambda)  \\ 
\mathcal{O}(\lambda) & \mathcal{O}(1) \end{array} \!\right],
\end{eqnarray}
from which we can infer
\begin{equation}
\begin{array}{ll}
Q_2 \sim \mathcal{O}(1)\,, \quad\quad\quad\quad\quad&Q_1 \sim  \mathcal{O}(\lambda)\,, \\ [3pt]
Q_{\dot{2}}^\dagger \sim \mathcal{O}(1)\,, &Q_{\dot{1}}^\dagger \sim  \mathcal{O}(\lambda)\,.
\end{array}
\label{eq:ChargeScaling}
\end{equation}
To leading power, only one supercharge ($Q_2$) is present in the EFT.
Expressing the supercharges as differential operators in superspace, and expanding on the light cone yields;
\be
\begin{array}{l}
\label{eq:SCETcharges}
Q_2 =  \left( i\, \frac{\partial}{\partial \theta^2} - \bar{\theta}^{ \dot{2}} \, \nbp - \sqrt{2} \,\bar{\theta}^{ \dot{1}}  \,\partial \right), \\[7pt]
Q_1 = \left( i \,\frac{\partial}{\partial \theta^1} -  \bar{\theta}^{\dot{1}}\, \np - \sqrt{2}\,  \bar{\theta}^{\dot{2}} \, \partial^*\right),
\end{array}
\ee
with analogous expressions for the conjugate charges.
Therefore the scaling of the momentum operator, and the supercharges as given in \eref{eq:ChargeScaling}, induce a non-trivial scaling of the superspace coordinates, see Table \ref{table:superspacescaling}.  Hence, two out of the four $\mathcal{N} = 1$ superspace Grassmann coordinates have a high virtuality and should not play a role in the EFT.

\begin{table}[t!]\vspace{0.1in}
\renewcommand{\arraystretch}{1.7}
\setlength{\tabcolsep}{0.35em}
\begin{center}
\begin{tabular}{| c || c | c | c | c |}
    \hline
    Coordinate &  $\theta^1 = \theta_2$ &  $\theta^{\dag \dot{1}} = \theta^{\dag}_{\dot{2}}$ &  $\theta^2 = - \theta_1$  & $\theta^{\dag \dot{2}} = - \theta^{\dag}_{ \dot{1}}$  \\ [2pt]\hline
    Scaling & $\lambda^{-1}$ & $\lambda^{-1}$ & $1$ & $1$  \\ \hline
\end{tabular}
\end{center}
\caption{Power counting for the superspace coordinates.}
\label{table:superspacescaling}
\end{table}

In terms of $y^\mu = x^\mu + i\, \theta \sigma^\mu \theta^{\dag}$, the superspace derivatives are $\bar{D}_{\dot{\alpha}} = - \partial/ \partial \theta^{\dag \dot{\alpha}}$.  Table \ref{table:superspacescaling} implies that they scale as $\bar{D}_{\dot{2}} \sim \mathcal{O}(1)$ and $\bar{D}_{\dot{1}} \sim \mathcal{O}(\lambda)$. To leading order in $\lambda$, $\{ D_2, \bar{D}_{\dot{2}} \} = - i \, \nbp  \sim \mathcal{O}(1)$, while all other components of the anti-commutator are suppressed.

Chiral and anti-chiral SCET superfields are defined such that they obey the EFT chirality condition, $D_2 \Phi^{\dagger} = 0 = \bar{D}_{\dot{2}} \Phi$. The physical degrees of freedom of the SCET LCG vector multiplet can be repackaged into a chiral superfield. Enforcing the chirality condition in the EFT, the chiral superfield $\Phi$ takes the form;
\bea
\Phi &=&  e^{-\frac{i}{2} \theta^{\dag \dot{2}} \theta^2\, \nbp} \left( \alc^{*}+ \theta^2 \,u_{n,2}^* \right) \nonumber \\  
&=&   \alc^{*} + \theta^2\, u_{n,2}^* -\frac{i}{2}\, \theta^{\dag \dot{2}} \,\theta^2 \,\nbp   \alc^{*}  \,,
\label{eq:SCETField}
\eea
where in the second line we have converted from $y^\mu$ to $x^\mu$ coordinates, dropped terms that are subleading in $\lambda$, and suppressed a gauge index in the case of non-Abelian fields.  There is only one complex fermionic degree of freedom in $\Phi$, and it obeys $P_n\, u_n^* = u_{n,2}^*$, since the spin up state has been projected out. Similarly, we have integrated out only one (spin-up) anti-collinear fermionic degree of freedom; this depends on the specific choice for $\bar{n}^\mu$. 

The (on-shell) SUSY transformations of the component fields in the EFT follow from the SCET expansion of the charges in \eref{eq:SCETcharges}.  Additionally, they are consistent with the expected component transformations of a chiral superfield: 
\begin{align}
&\delta_\eta u_{n,2} = i\, \sqrt{2}\, \eta^{\dagger \dot{2}}\, \nbp  \alc \,, &\delta_\eta  \alc = \sqrt{2}\, \eta^2 \,u_{n,2} \,,
\label{eq:SUSYTrans}
\end{align}
where we have used $\left(n\cdot \sigma\right)_{2 \dot{2}} = 2$.
The collinear SCET Lagrangian is invariant under these transformations~\cite{SUSYSCET}.

Now that we have explored some general aspects of marrying SCET and SUSY, we will focus our attention on a specific example.
In the rest of this letter, we will apply the general algorithm presented above to the free Abelian gauge theory.  Then we will conclude by quoting the result for non-Abelian gauge theory~\cite{SUSYSCET}.

Since SUSY is a good symmetry, the projection operators acting on the gauginos of a vector multiplet imply that the entire superfield obeys the decomposition:
\bea
\label{eq:SuperfieldProjections}
V = V^\dagger = P_n\,V+P_{\bar{n}}\,V =  V_n + V_{\bar{n}}\,,
\eea 
where the projection operators are defined in \eref{eq:projectionOps}.  Using $u_{n,1}= 0 =u_{\bar{n},2}$,  the collinear and anti-collinear on-shell superfields are
\begin{widetext}
\bea
V_n  &=& - \theta^1\, \theta^{\dag \dot{1}}\, n \cdot A_n - \sqrt{2}\, \Big(\theta^1\,\theta^{\dag \dot{2}}  \alc_n^* + \theta^2\, \theta^{\dag \dot{1}} \, \alc_n\Big) + 2\,i\, \theta^1\, \theta^2\,  \theta^{\dag \dot{2}}\, u^*_{n, \dot{2}} - 2\,i\, \theta^{\dag \dot{1}}\, \theta^{\dag \dot{2}}\, \theta^2\, u_{n,2}\,,   \nonumber \\[2pt]
V_{\bar{n}}   &=& -\theta^1\, \theta^{\dag\dot{1}}\, n \cdot A_{\bar{n}} 
-\theta^2 \,\theta^{\dag\dot{2}}\, \bar{n} \cdot A_{\bar{n}} - \sqrt{2}\,\Big( \theta^1 \,\theta^{\dag\dot{2}} \, \alc^*_{\bar{n}}  + \theta^2 \,\theta^{\dag\dot{1}}\,  \alc_{\bar{n}}\Big)
+ 2\,i\, \theta^1\, \theta^2 \,\theta^{\dag\dot{1}}\,u^*_{\bar{n}, \dot{1}}  - 2\,i\, \theta^{\dag\dot{1}}\,\theta^{\dag\dot{2}}\, \theta^1\, u_{\bar{n},1}   \,,
\label{eq:VectorSuperfields}
\eea 
\end{widetext}
where we have fixed the LCG condition $\bar{n}\cdot A_n = 0$.

The action for the Abelian theory is 
\bea
S = \int \text{d}^4 x \,\text{d}^2 \theta\,  \mathcal{W}^\alpha \mathcal{W}_\alpha + \text{h.c.}\,, 
\label{eq:FullAction}
\eea
where $\mathcal{W}_\alpha$ is a chiral superfield which in Wess-Zumino gauge is
\bea
\mathcal{W}_\alpha =-\frac{i}{4}\, \bar{D}\bar{D}\,D_\alpha \big(V_n + V_{\bar{n}}\big)\,,
\eea
where $DD = D^\alpha D_\alpha$ and $\bar{D}\bar{D} = \bar{D}_{\dot{\alpha}} \bar{D}^{\dot{\alpha}}$.

The anti-collinear vector superfield can be integrated out using the variation of the superspace action. 
This yields a superspace constraint equation, $D^\alpha\mathcal{W}_\alpha = 0$, which encodes the equation of motion for $V_{\bar{n}}$;
\begin{align}
\Big(-\!16\, \Box + 4\,i\, D^\alpha \big(\sigma \cdot \partial\big)_{\alpha \dot{\alpha}}\, \bar{D}^{\dot{\alpha}} \Big) \big(V_n + V_{\bar{n}}\big) = 0 \,.
\label{eq:VectorSuperfieldConstraint}
\end{align}
It is instructive to see that the equations of motion for the component fields that are integrated out in the EFT, $u_{\bar{n}}$ and $n\cdot A_n$, are equivalent to this constraint equation.  To isolate the leading order fermionic components of the vector superfield expanded in \eref{eq:VectorSuperfields}, apply $\bar{D}_{\dot{2}}$ to the constraint equation: \\
\begin{align}
\bar{D}_{\dot{2}}\, D^2\, \big(\sigma \cdot \partial\big)_{2 \dot{2}} \,\bar{D}^{\dot{2}}\,V_{\bar{n}} &= - \bar{D}_{\dot{2}} \,D^1 \,\big(\sigma \cdot \partial\big)_{1 \dot{2}} \,\bar{D}^{\dot{2}}\,V_n\,;  \notag \\[2pt]
\Longrightarrow \quad u_{\bar{n}, 1} &= \frac{\sqrt{2}\, \partial^*}{\nbp} \,u_{n,2}\,,
\label{eq:EOMgaugino}
\end{align}
which reproduces the expected equation of motion for the anti-collinear gaugino, see~\eref{eq:anticollEOM}.
Additionally, it is straightforward to show that \eref{eq:VectorSuperfieldConstraint} integrates out the unphysical gauge polarization $n\cdot A_n$, thereby reproducing the LCG Lagrangian.  

This motivates an ansatz for the equation of motion of the anti-collinear vector superfield:
\begin{align}
V_{\bar{n}} = - V_n &-  \frac{1}{\nbp D_2 \,\bar{D}_1 \, D_1   }  \bar{D}_{\dot{2}} \,DD \,\Big(\bar{D}_{\dot{2}}\, D_1\,  V_n\Big) \notag \\
& - \frac{1}{\nbp  \,  \bar{D}_{\dot{2}}  \,D_1 \, \bar{D}_{\dot{1}}}  D_2\, \bar{D} \bar{D}\, \Big(D_2\, \bar{D}_{\dot{1}}\, V_n\Big) \,.
\label{eq:VectorAnsatz}
\end{align}
Both nontrivial terms are required to ensure the reality condition $V_{\bar{n}} = V_{\bar{n}}^\dagger$.  Dividing by superspace derivatives is well-defined by taking a super-Fourier transform and considering momentum and super-momentum eigenvalues. 
\eref{eq:VectorAnsatz} satisfies the constraint equation \eref{eq:VectorSuperfieldConstraint}.  Furthermore, it reproduces the component equations of motion for unphysical degrees of freedom.  For example, projecting with $D_2  \bar{D}_1D_1\,$ reproduces \eref{eq:EOMgaugino}. 

In the LCG EFT the remaining physical degrees of freedom $u_n$ and $\alc$ form a chiral superfield. This can be justified in superspace by taking projections on a vector superfield \eref{eq:VectorSuperfields}, for instance
\bea
\nonumber
 \Phi \equiv \bar{D}_{\dot{2}}\, D_1\, V_n \Big|_{\theta^1 = 0 = \theta^{\dagger \dot{1}}} && \\ 
  = \sqrt{2} \,\alc^*\! +2\,i\, \theta^2 u_{n,2}^* &+&  i\, \sqrt{2}\, \theta^2 \theta^{\dagger \dot{2}} \,\nbp \alc^*, 
\label{eq:ChiralFromVector}
\eea
which obeys the chirality constraint $\bar{D}_{\dot{2}}\, \Phi = 0$; for the anti-chiral multiplet, simply take the conjugate of \eref{eq:ChiralFromVector}. Therefore, the ansatz for integrating out the anti-collinear modes \eref{eq:VectorAnsatz} can be expressed in terms of the chiral and anti-chiral superfields. 

After some manipulations, the action \eref{eq:FullAction} is $S \propto \int \text{d}^4 x\, \text{d}^4 \theta\,  \bar{D}_{\dot{1}}\, D^\alpha\, \big(V_n + V_{\bar{n}}\big)\,  \bar{D}_{\dot{2}} \,D_\alpha \, \big(V_n + V_{\bar{n}}\big) $. Using \eref{eq:VectorAnsatz} to integrate out the anti-collinear superfield yields the EFT action\footnote{Recall that $\int \text{d} \theta^\alpha D_\alpha (\textrm{. . . })$ is a total derivative in real space, and therefore we can drop surface terms when using integration by parts if we assume that they vanish sufficiently fast at infinity. In SCET, integration by parts is well defined for the inverse derivative operator $1/\nbp$ because it can be cast in terms of its momentum space representation. By analogy we extend this argument and use integration by parts on $1/D$ operators in the following calculation.}
\begin{widetext}
\bea
\nonumber
\mathcal{L} &=& \int \text{d}^4 \theta  \left( \frac{1}{\nbp \, \bar{D}_{\dot{1}} } D^1\, \bar{D} \,\bar{D} \, \Big(D_2 \bar{D}_{\dot{1}} V_n\Big) \right) \left(\frac{1}{\nbp \,D_1}  \bar{D}^{\dot{1}}\,D\,D \,\Big(\bar{D}_{\dot{2}} D_1 V_n \Big)   \right)   \\ \nonumber
&=&\int \, \text{d}\theta^2 \,\text{d} \theta^{\dagger \dot{2}} \,   \text{d} \theta^{\dagger \dot{1}}\, \text{d} \theta^1  \, \frac{1}{ \,D_1  \, \bar{D}_{\dot{1}}} \Big( D^1 \,\bar{D} \bar{D} \, \Big(D_2 \bar{D}_{\dot{1}} V_n\Big)\Big) \frac{1}{(\nbp)^2} \left( \bar{D}^{ \dot{1}} \,DD \,\Big(\bar{D}_{\dot{2}} D_1 V_n \Big) \right) \\ 
&=& \int \text{d}\theta^2 \, \text{d} \theta^{\dagger \dot{2}} \, \Phi_{n}^\dagger \, \frac{ \bar{D}\bar{D} D_2 \bar{D}_{\dot{2}} DD }{\left( \nbp \right) ^2}\Phi_{n} = \int  \text{d}\theta^2\, \text{d} \theta^{\dagger \dot{2}}  \,\Phi_{n}^\dagger \frac{i \,\Box}{\nbp}\Phi_n  \quad \subset\quad  i\, u_{n,2}^* \left( n \cdot \partial + \frac{\partial_{\perp}^2}{\nbp} \right) u_{n,2} + \alc^* \Box \alc \, ,
\eea
\end{widetext}
which reproduces the expected equation of motion in the free theory. 
We conclude that integrating out the anti-collinear fermion translates into integrating out two superspace coordinates, namely $\theta^1\sim 1/\lambda$ and $\theta^{\dagger {1}} \sim 1/\lambda$, while $\theta^2\sim 1$ and $\theta^{\dagger {2}} \sim 1$ remain in the EFT. Note that in the above calculation we can identify the various projections of $V_n$ with a chiral superfield by \eref{eq:ChiralFromVector} in the EFT.

Finally for completeness, we quote the result for the collinear superspace LCG Lagrangian in $\mathcal{N}=1$ SYM.  This model is invariant under the SUSY transformations \eref{eq:SUSYTrans} and meets additional requirements such as RPI demonstrating that it is a consistent collinear EFT~\cite{SUSYSCET}:
\begin{widetext}
\be
\label{eq:VectorSuperspaceLagrangian}
\mathcal{L} = \int \!\text{d} \theta^2 \text{d} \theta^{\dagger \dot{2}}\bigg[\Phi^{\dagger a} \frac{\Box}{\nbp} \Phi^a 
+ 2\, g \left( f^{abc}\,  \Phi^a\,  \Phi^{\dagger b} \frac{\partial^\star}{\nbp} \Phi^c + \text{h.c.} \right) 
+ 2 \,g^2 \, f^{abc}f^{ade} \frac{1}{\nbp} \Big( \Phi^b \bar{D}_{ \dot{2}}  \Phi^{\dagger c}  \Big) \frac{1}{\nbp} \Big(  \Phi^{\dagger d} D_2  \Phi^e \Big)\bigg].
\vspace{20pt}
\ee
\end{widetext}

Note that the form of this expression is what one would have naively obtained by supersymmetrizing the pure LCG Yang Mills Lagrangian \cite{Leibbrandt:1987qv}. In this sense, the  on-shell collinear EFT makes SUSY transparent. While similar expressions to \eref{eq:VectorSuperspaceLagrangian} do exist in the literature for $\mathcal{N} = 4$ SYM \cite{Mandelstam:1982cb,Brink:1982pd}, the present work simultaneously provides the first application to collinear fields along with a general algorithm that can be used to derive the Lagrangian. 

In conclusion, this letter has provided a framework for studying SUSY in the collinear limit.  A general algorithm for deriving an EFT defined on collinear superspace was proposed, and it was applied to the case of an $\mathcal{N} = 1$ Abelian superfield.  We also provided the result for a non-Abelian theory.  In a followup work~\cite{SUSYSCET}, we will provide a more complete treatment of the EFT perspective, including a detailed discussion of the remaining symmetries of the EFT, and an explanation of how the Super-Poincare generators reduce to RPI.  This will provide the groundwork for many interesting extensions, including models with a larger number of supercharges, and even perhaps theories of collinear supergravity. 

\vspace{-0.26in}
\section*{Acknowledgments}
\vspace{-0.1in}
We are grateful to Marat Freytsis and Duff Neill and for helpful comments.  TC is supported by an LHC Theory Initiative Postdoctoral Fellowship, under the National Science Foundation grant PHY--0969510.  GE is supported by the U.S. Department of Energy, under grant Contract Numbers DE--SC00012567.  AL is supported an LHC Theory Initiative Postdoctoral Fellowship, under the National Science Foundation grant PHY--1419008. This work was in part initiated at the Aspen Center for Physics, which is supported by National Science Foundation grant PHY--1066293.

\end{spacing}

\bibliography{SUSY_SCET}
\bibliographystyle{utphys}

\end{document}